\documentclass[11pt,a4]{article}
\usepackage{graphicx} %
\usepackage{amsmath}
\usepackage{lineno}

\usepackage{relsize}
\usepackage{graphicx}
\usepackage{tcolorbox}
\usepackage{todonotes}
\newcommand{\ignore}[1]{{}}
\usepackage[colorlinks=true,
allcolors=blue,
            pdfborder={0 0 0},
            ]{hyperref}
\date{November 13, 2025}

\title{Update to the U.S. National Input to\\the European Strategy Update for Particle Physics}
\author{Andr\'e de Gouv\^ea, Hitoshi Murayama, Mark Palmer, Heidi Schellman for the\\ Executive Committees of the Division of Particles and Fields and Division of Physics of Beams\\ of the American Physical Society}

\begin{document}

\maketitle
\newcommand{\Snowmass}{\href{https://inspirehep.net/literature/2623768}{2021 Snowmass Community Planning Exercise}}

\newcommand{\Pfivereport}{\href{https://www.usparticlephysics.org/2023-p5-report/}{P5 report} }

\newcommand{\Pfive}{\href{https://www.usparticlephysics.org/2023-p5-report/}{P5} }

\newcommand{\nasemreport}{\href{https://nap.nationalacademies.org/catalog/28839/elementary-particle-physics-the-higgs-and-beyond}{Elementary Particle Physics: The Higgs and Beyond}}

\newcommand{\nasem}{\href{https://nap.nationalacademies.org/catalog/28839/elementary-particle-physics-the-higgs-and-beyond}{NASEM}}

\newcommand{\nasemrecommendations}{\href{https://nap.nationalacademies.org/catalog/28839/elementary-particle-physics-the-higgs-and-beyond}{recommendations}}

\newcommand{\hfccpage}{\href{https://higgsfactory.slac.stanford.edu/about}{U.S. Higgs Factory Circular Collider} }

\newcommand{\hfccsubmission}{\href{https://indico.cern.ch/event/1439855/contributions/6461597/}{submission}}

\newcommand{\hfccinput}{\href{https://indico.cern.ch/event/1439855/contributions/6461597/}{input}}

\newcommand{\usmccpage}{\href{https://www.muoncollider.us}{U.S. Muon Collider Collaboration} }

\newcommand{\usmccsubmission}{\href{https://indico.cern.ch/event/1439855/contributions/6461576/}{submission}}

\newcommand{\usmccstudy}{\href{https://www.muoncollider.us/otherorgs/nationallabstudygroup/}{ National Lab Accelerator Study Group for a Muon Collider}}

\newcommand{\magnetreport}{\href{https://arxiv.org/pdf/2508.19220}{The 2025 Roadmaps for the US Magnet Development Program}}

\newcommand{\dpfsubmission}{\href{https://agenda.infn.it/event/44943/contributions/263481/}{reported}}
\newcommand{\dpfdoc}[1]{\href{https://indico.cern.ch/event/1439855/contributions/6461648/}{#1}}

\thispagestyle{empty}

\setlength\itemsep{-3em}
\begin{abstract}
   In this document we update the status of U.S. community inputs for the European Strategy for Particle Physics Update (ESPPU) since April 1, 2025, and offer responses to the revised questions. Major new inputs include a long-term strategy report from the National Academies of Sciences, Engineering, and Medicine  and the formal formation of a U.S. Muon Collider Collaboration.
    \end{abstract}

\subsubsection*{The U.S. planning process and context}

This document is an update of our previous \dpfdoc{submission} \cite{deGouvea:2025lfx}, that reported on the outcome of the U.S. long term planning process.  That process started with the  \Snowmass~\cite{Butler:2023eah}, organized by the Divisions of Particles and Fields and the Physics of Beams (DPF and DPB) of the American Physical Society.  

That broad community process was input to a focused prioritization process in which a Particle Physics Project Prioritization Panel (P5) was charged by the National Science Foundation (NSF) and the Department of Energy (DOE) to develop a 10-year strategic plan for U.S. particle physics, in the context of a 20-year global strategy and two constrained budget scenarios.  The \Pfivereport~\cite{P5:2023wyd}  was approved by the federal High Energy Physics Advisory Panel (HEPAP), in December, 2023. 3,200 U.S. scientists signed an expression of support for the report.  

For the Venice meeting, DPF \dpfsubmission~\cite{deGouvea:2025lfx,HitoshiVenice} on the P5 process and the first iteration of a Higgs Factory \href{https://indico.cern.ch/event/1439855/contributions/6461597/}{study}. Since then, the U.S. National Academies of Sciences, Engineering and Medicine (NASEM) has produced a report, \nasemreport~\cite{NationalAcademiesofSciencesEngineeringandMedicine:2025bix}.

\subsubsection*{The National Academies Report}

In parallel with the P5 process, and sharing many of the same community inputs,  the National Academy of Sciences Committee on Elementary Particle Physics was tasked by the DOE and the NSF
 with setting a long-term vision for the field, focusing on innovation and new approaches.  Their \href{https://nap.nationalacademies.org/catalog/28839/elementary-particle-physics-the-higgs-and-beyond}{report}~\cite{NationalAcademiesofSciencesEngineeringandMedicine:2025bix} was released in June 2025. %
Where the \Pfivereport focused on strategy for the next 10-20 years, the \nasem\ report had a 40-year outlook. 
The relevant NASEM recommendations are reproduced below.

\begin{quotation}

{\noindent \bf Recommendation 1:\ }\hypertarget{nas1} The United States should host the world's highest-energy elementary
particle collider around the middle of the century. This requires the immediate creation of a
national muon collider research and development program to enable the construction of a
demonstrator of the key new technologies and their integration.

{\noindent\bf Recommendation 2:\ }\hypertarget{nas2} The United States should participate in the international Future
Circular Collider Higgs factory currently under study at CERN to unravel the physics of
the Higgs boson.

{\noindent\bf Recommendation 3:\ }\hypertarget{nas3} The United States should continue to pursue and develop new
approaches to questions ranging from neutrino physics and tests of fundamental
symmetries to the mysteries of dark matter, dark energy, cosmic inflation, and the excess of
matter over antimatter in the universe.

...

{\noindent\bf Recommendation 5:\ }\hypertarget{nas5} The United States should invest for the long journey ahead with
sustained research and development funding in accelerator science and technology,
advanced instrumentation, all aspects of computing, emerging technologies from other
disciplines, and a healthy core research program.

...

{\noindent\bf Recommendation 7:}\hypertarget{nas7} The United States should engage internationally through existing and
new partnerships and explore new cooperative planning mechanisms.

\end{quotation}

These recommendations are similar to the P5 recommendations but further emphasize the priority of strong participation in the next collider at CERN and development of a US-based muon collider.

\vspace{-1 em}
\subsubsection*{Additional developments since the Venice meeting}
\begin{itemize}
\setlength\itemsep{0em}
    \item The U.S. DOE Office of Science has repurposed the   Higgs Factory Coordination Consortium  (HFCC) activity as a \href{https://higgsfactory.slac.stanford.edu/about}{U.S. Higgs Factory Circular Collider} organization  concentrating on the FCC-ee proposal. 
       It remains charged to provide strategic direction and leadership for the U.S. community to engage, shape, and thereby advance the development of the physics, experiment, and detector (PED) and accelerator (A) programs for a potential future Higgs factory; now with greater emphasis on FCC-ee. %

    \item The \href{https://www.muoncollider.us}{US Muon Collider Collaboration} (USMCC) has been formalized in the U.S. to coordinate national muon collider activities and work closely with the \href{https://muoncollider.web.cern.ch}{International Muon Collider Collaboration}.%
    
        \item In consultation with the \href{https://www.muoncollider.us}{USMCC}, the directors of  eight U.S. national laboratories with accelerator programs created a short-term \href{https://www.muoncollider.us/otherorgs/nationallabstudygroup/}{ National Lab Accelerator Study Group for a Muon Collider} to evaluate the pressing needs for muon collider R\&D and assess how each laboratory’s strengths can contribute to the effort. This group is charged to provide an independent assessment of the progress and future of the program, with a final report expected in mid-2026.

    \item The U.S. Magnet R\&D community has produced a new report \href{https://arxiv.org/pdf/2508.19220}{The 2025 Roadmaps for the U.S. Magnet Development Program}~\cite{Cooley:2025gtt}.

\end{itemize}

The current budget for particle physics in the U.S. is unfortunately below the level of  the less optimistic P5 funding scenario.  NSF funding, which supports detectors, computing and individual investigators, is particularly impacted.

\subsection*{Response to {ECFA questions}}
{%

{\bf 1. What is your preferred large-scale post-LHC accelerator for CERN?}

 Both the \Pfive and \nasem\ reports agree that the preferred next collider should be a Higgs factory. P5 concluded that such a facility could not be hosted by the U.S. within the budget guidelines from the agencies. P5 identified both FCC-ee and a linear collider as projects that could meet the scientific requirements without specifying their locations. The \nasem\ report more explicitly recommends (\hyperlink{nas2}{recommendation 2}) U.S. participation in the international Future Circular Collider Higgs factory currently under study at CERN.

The U.S \href{https://higgsfactory.slac.stanford.edu/about}{Higgs Factory Coordination Consortium} submitted their \hfccinput\ to the European Strategy in March.
\begin{quotation}
\noindent
    ``The U.S. is enthusiastic for a Higgs Factory as the next major collider and strongly supports FCC-ee, intending to collaborate on its construction and physics exploitation if it is chosen as the next major research infrastructure project at CERN. ... The U.S.  would also support an LC if the CERN Council approves such a project in a timely manner. The U.S. eagerly awaits a CERN Council decision and looks forward to partnering with CERN on the next future collider project.''\cite{hfccsubmission}
\end{quotation}

The  DOE has since supported focusing U.S. studies on  FCC-ee detector and machine development.

{\noindent\bf 2. What is your preferred alternative, if the preferred option would not be feasible?} 

The U.S. community is enthusiastic about participation in CERN's next collider.    As noted above, there is now an official U.S. Higgs Factory Circular Collider working group directed towards the FCC-ee option. If the FCC-ee is not feasible, there are substantial communities exploring a Linear Collider Facility \cite{lcvision}, as well as alternate paths to high-energy colliders \cite{usmccsubmission,lep3,clic,wakefield,lowenergyhh,fcchh}. The specific choice would depend on the reasons why the preferred option is not feasible and the timescale to physics of the alternatives. A more definitive answer awaits further input from the U.S. community and the ESPPU. 

{\noindent\bf 3. What is your preferred alternative, if the preferred option would not be competitive? 
}        

 Answering this question requires a better understanding of global plans for large projects than we currently have. P5 recommended that the U.S. should revisit this question, in light of international developments, later in this decade.

\subsubsection*{Beyond the next collider}

The USMCC has engaged a vibrant and growing community that is focused on critical R\&D and a Muon Collider Demonstrator to support eventual construction of a Muon Collider.
The new \nasem\ report (\hyperlink{nas1}{ recommendation 1}) states
"The United States should host the world's highest-energy elementary
particle collider around the middle of the century. This requires the immediate creation of a
national muon collider research and development program to enable the construction of a
demonstrator of the key new technologies and their integration." %

From the USMCC \usmccsubmission\ to the \href{https://europeanstrategyupdate.web.cern.ch}{European Strategy for Particle Physics Update}: 
\begin{quotation}
``Assuming expanded funding and successful technological developments, we anticipate achieving
technical design readiness for the collider within approximately 20 years, which could pave the way for the
start of a construction project in the mid-2040s and operations commencing soon thereafter.''\cite{usmccsubmission}
\end{quotation}

\subsubsection*{The broader U.S. research program}
The next collider is not the only priority for the U.S. program. Both the \Pfive and \nasem\ reports emphasized the need for a broader program; both to exploit existing construction projects such as the HL-LHC and DUNE (P5 recommendation 1) \cite{P5:2023wyd} and ``to pursue and develop new approaches to questions ranging from neutrino physics and tests of fundamental symmetries to the mysteries of dark matter, dark energy, cosmic inflation, and the excess of matter over antimatter in the universe.'' (\hyperlink{nas3}{NASEM recommendation 3}).  These efforts, wherever they are located,  rely on  mutual international collaboration, including essential contributions from CERN and European institutions.

In parallel, we must build further upon our global R\&D collaborations spanning computation, theory, and technology development that will provide the foundation for the future of the field.

\vspace{-1 em}
\subsection*{Conclusion}
In this document, we have summarized recent updates in the U.S. community prioritization process.  The \Pfive and \nasem\ reports are the result of broad community input followed by a rigorous prioritization process, only retaining the most compelling projects that are consistent with known budget and time constraints.  The P5  priorities have been explicitly endorsed by a majority of the U.S. particle physics community and reinforced by the \nasem\ report.

US and European particle physics efforts have benefited tremendously from mutual engagement.  The U.S. community and funding agencies continue to pursue collaboration towards the FCC-ee at CERN and a Muon Collider, potentially hosted by the U.S. %
While CERN priorities remain a predominantly European decision, our mutual goal must be worldwide optimization of resources for high-impact science.

}


\begin{thebibliography}{20}
\newcommand{\esg}{ Input to the European Strategy for Particle Physics Update,
April 2025.}

\bibitem{deGouvea:2025lfx}
A.~de Gouv{\^e}a, H.~Murayama, M.~Palmer and H.~Schellman,
``US National Input to the European Strategy Update for Particle Physics,''
[\href{https://arxiv.org/abs/2504.01804}{arXiv:2504.01804} [hep-ex]], April 1, 2025.


\bibitem{Butler:2023eah}
J.~N.~Butler, R.~S.~Chivukula, A.~de Gouv{\^e}a, T.~Han, Y.~K.~Kim, P.~Cushman, G.~R.~Farrar, Y.~G.~Kolomensky, S.~Nagaitsev and N.~Yunes, \textit{et al.}
``Report of the 2021 U.S. Community Study on the Future of Particle Physics (Snowmass 2021) Summary Chapter,''
[\href{https://arxiv.org/abs/2301.06581}{arXiv:2301.06581} [hep-ex]].

\bibitem{HitoshiVenice}
H.~Murayama, ``\href{https://agenda.infn.it/event/44943/contributions/263481/}{Large-scale particle physics projects in the US and plans for participation in projects outside}", presentation at the Open Symposium on the European Strategy for Particle Physics, June 24, 2025.


\bibitem{P5:2023wyd}
S.~Asai \textit{et al.} [P5],
``Exploring the Quantum Universe: Pathways to Innovation and Discovery in Particle Physics,''
\href{https://doi.org/10.2172/2368847}{doi:10.2172/2368847}
[\href{https://arxiv.org/abs/2407.19176}{arXiv:2407.19176} [hep-ex]].


\bibitem{NationalAcademiesofSciencesEngineeringandMedicine:2025bix}
National~Academies~of~Sciences, Engineering, and Medicine,
``Elementary Particle Physics: The Higgs and Beyond,''
The National Academies Press, 2025,
ISBN 978-0-309-99333-3, 978-0-309-73277-2
\href{https://doi.org/10.17226/28839}{doi:10.17226/28839}

\bibitem{hfccsubmission}
[HFCC Collaboration],``\href{https://indico.cern.ch/event/1439855/contributions/6461597/}{U.S. Higgs Factory Consortium input to ESG on CERN's future collider options}", April 1, 2025.

\bibitem{usmccsubmission}
[USMCC Collaboration]``\href{https://indico.cern.ch/event/1439855/contributions/6461576/}{United States Muon Collider Community White Paper for the European Strategy for Particle Physics Update}", April 1, 2025.

\bibitem{Cooley:2025gtt}
L.~Cooley, P.~Ferracin, S.~Gourlay, D.~Larbalestier, M.~Palmer, S.~Prestemon, G.~Velev, G.~Ambrosio, D.~Arbelaez and K.~Badgley, \textit{et al.}
``The 2025 Roadmaps for the US Magnet Development Program,
[\href{https://arxiv.org/abs/2508.19220}{arXiv:2508.19220} [physics.acc-ph]].

\bibitem{lcvision}
``\href{https://arxiv.org/abs/2503.19983}{A Linear Collider Vision for the Future of Particle Physics}'', \esg

\bibitem{lep3}
``\href{https://indico.cern.ch/event/1439855/contributions/6461601/}{LEP3: A High-Luminosity $e^+e^-$ Higgs \& Electroweak Factory in the LHC Tunnel}'', \esg

\bibitem{clic}
``\href{https://indico.cern.ch/event/1439855/contributions/6461475/}{The Compact Linear $e^+e^-$ Collider (CLIC)}'', \esg

\bibitem{wakefield}
``\href{https://www.overleaf.com/project/68ec47790b8d783d9f9527e2}{Design Initiative for a 10 TeV pCM Wakefield Collider}'', \esg

\bibitem{lowenergyhh}
``\href{https://indico.cern.ch/event/1439855/contributions/6461643/}{Physics Prospects for a near-term Proton-Proton Collider}'', \esg

\bibitem{fcchh}
``\href{https://indico.cern.ch//event/1439855/contributions/6461636/}{FCC Integrated Programme Stage 1: The FCC-ee},
\href{https://indico.cern.ch//event/1439855/contributions/6461658/}{FCC Integrated Programme Stage 2: The FCC-hh}'', \esg

\bibitem{lehc}
``\href{https://indico.cern.ch/event/1439855/contributions/6461469/}{Future Opportunities with Lepton-Hadron Collisions}", \esg

\end{thebibliography}
\end{document}